\documentclass{cjaa}
\usepackage{graphicx}                   
\input{epsf.sty}                        
\input{psfig.sty}                       
\begin{document}
\title{Radio Spectra of three Supernova Remnants: G114.3+0.3, G116.5+1.1 and G116.9+0.2}
\volnopage{Vol.0 (200x) No.0, 000--000}      
\setcounter{page}{1}          

\author{Wenwu Tian\inst{1,2}  
 \and
        Denis Leahy\inst{2}}
\authorrunning{W.W. Tian$\&$D. Leahy} 
\offprints{W.W. Tian and D. Leahy}
\institute{National Astronomical Observatories, CAS, Beijing 100012, China\\
\and
Department of Physics \& Astronomy, University of Calgary, Calgary, Alberta T2N 1N4, Canada}
  
\date{Received Dec. 18, 2005; accepted Jan., 2006} 

\abstract{
New images of the Supernova Remnants (SNRs) G114.3+0.3, G116.5+1.1 and G116.9+0.2 (CTB 1) 
are presented at 408 MHz from the Canadian Galactic Plane Survey (CGPS). We also use the
1420 MHz images from the CGPS to study their 408-1420 MHz spectral indices.
The flux densities at 408 MHz and 1420 MHz, 
corrected for flux densities from compact sources within the SNRs, are 12$\pm$6 Jy and 9.8$\pm$0.8 Jy for G114.3+0.3, 15.0$\pm$1.5 Jy and 10.6$\pm$0.6 Jy for G116.5+1.1, 15.0$\pm$1.5 Jy and 8.1$\pm$0.4 Jy for G116.9+0.2.  
The integrated flux density-based spectral indices (S$_{\nu}$$\propto$$\nu$$^{-\alpha}$)  
are $\alpha$=0.16$\pm$0.41, 0.28$\pm$0.09 and 0.49$\pm$0.09 for G114.3+0.3, G116.5+1.1 and G116.9+0.2, respectively. Their T-T plot-based spectral indexes are 0.68$\pm$0.48, 0.28$\pm$0.15, 
and 0.48$\pm$0.04, in agreement with the integrated flux density-based spectral indices. 
New flux densities are derived at 2695 MHz  
which are significantly larger than previous values. 
The new 408, 1420 and 2695 MHz flux densities and
published values at other frequencies, where images are not available, are fitted
after correcting for contributions from compact sources, to derive their multi-frequency spectral indices. 
\keywords{ISM:individual G114.3+0.3, G116.5+1.1, G116.9+0.2 - radio continuum:ISM}}

\titlerunning{Radio Spectrum of three SNRs}
\maketitle 

\section{Introduction}
\label{sect:Intro}
As one of the systematric efforts to study supernova remnants spectral indice variation, we report new CGPS radio observations of three shell-type SNRs, i.e. G114.3+0.3, G116.5+1.1 and G116.9+0.2 (CTB 1) which are within a field of 4$^{o}$$\times$ 2.5$^{o}$. Their radio spectra have not been studied in detail before. G116.9+0.2 (CTB 1) has been extensively studied in both radio (e.g. Yar-Uyaniker et al., 2004) and optical emission (e.g. Fesen et al., 1997).
G114.3+0.3 and G116.5+1.1, due to their low surface brightness and large diameter, 
are less well studied (Reich $\&$ Braunsfurth, 1981; Yar-Uyaniker et al., 2004; Mavromatakis et al. 2005).  
Reich $\&$ Braunsfurth (1981) gave their images and physical properties at 1420 MHz with 
HPBW 9$\times$9 arcmin and 2696 MHz with 4.6$\times$4.6 arcmin using the Effelsberg 100-m telescope. 
Yar-Uyaniker et al. (2004) presented their continuum and HI images at 1420 MHz with HPBW 
about 1 arcmin from the Canadian Galactic Plane Survey (CGPS), mainly emphasizing distance estimates.
Mavromatakis et al. (2005) present optical and HI observations of G116.5+1.1.

In this paper, we use both 408 MHz and 1420 MHz data from CGPS, and 2695 MHz data from the 
Effelsberg 100-m telescope (Fu\"rst et al., 1990) in order to study the three SNRs' radio spectra.    

\section{Observations and Image Analysis}
\label{sect:Obs}
The 408 MHz and 1420 MHz data sets come from the CGPS,
which is described in detail by Taylor et al. (2003).
The data sets are mainly based on observations from the Synthesis Telescope 
(ST) of the Dominion Radio Astrophysical Observatory (DRAO). The spatial
resolution is better than 1'$\times$ 1' cosec($\delta$) 
at 1420 MHz and 3.4'$\times$3.4' cosec($\delta$) at 408 MHz.  
DRAO ST observations 
are not sensitive to structures larger than an angular 
size scale of about 3.3$^{o}$ at 408 MH and 56' at 1420 MHz. Thus the CGPS includes 
data from the 408 MHz all-sky survey of Haslam et al (1982) which has an 
effective resolution of 51' and the Effelsberg 1.4 GHz Galactic plane survey 
of Reich et al. (1990, 1997) with resolution 9.4' for large scale emission 
(the single-dish data are freely available at http://www.mpifr-bonn.mpg.de/survey.html). 
 
We analyze the images and determine flux
densities using the DRAO export software package.  
Two of our previous papers (Tian and Leahy, 2005; Leahy and Tian, 2005) have given a detailed description of the methods to reduce the influence of compact sources within the SNRs on the radio spectrum.
 
\section{Results}
\label{sect:Res}

\begin{figure*}
\vspace{90mm}
\begin{picture}(300,350)
\put(0,360){\includegraphics{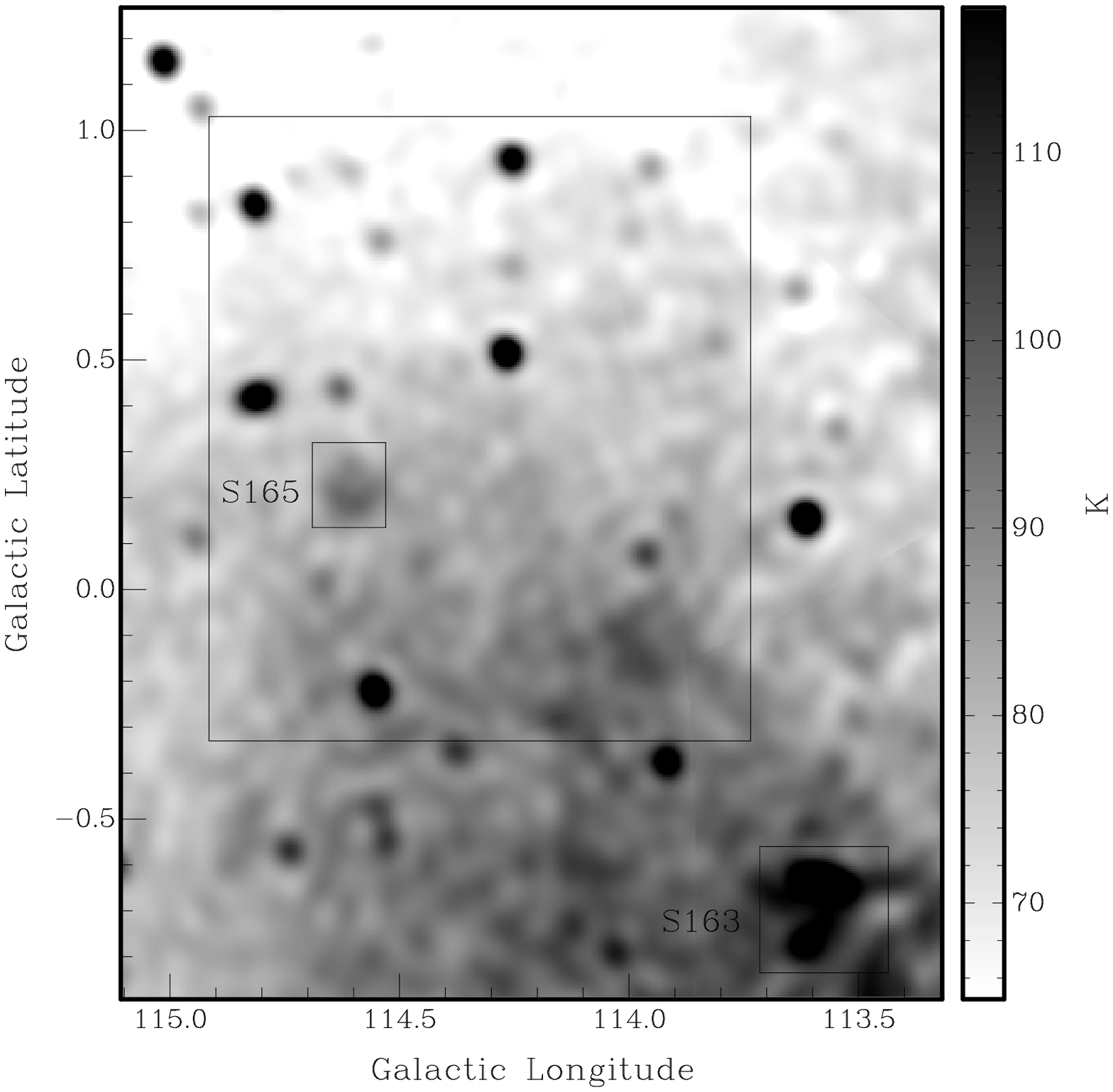}}
\put(208,350){\includegraphics{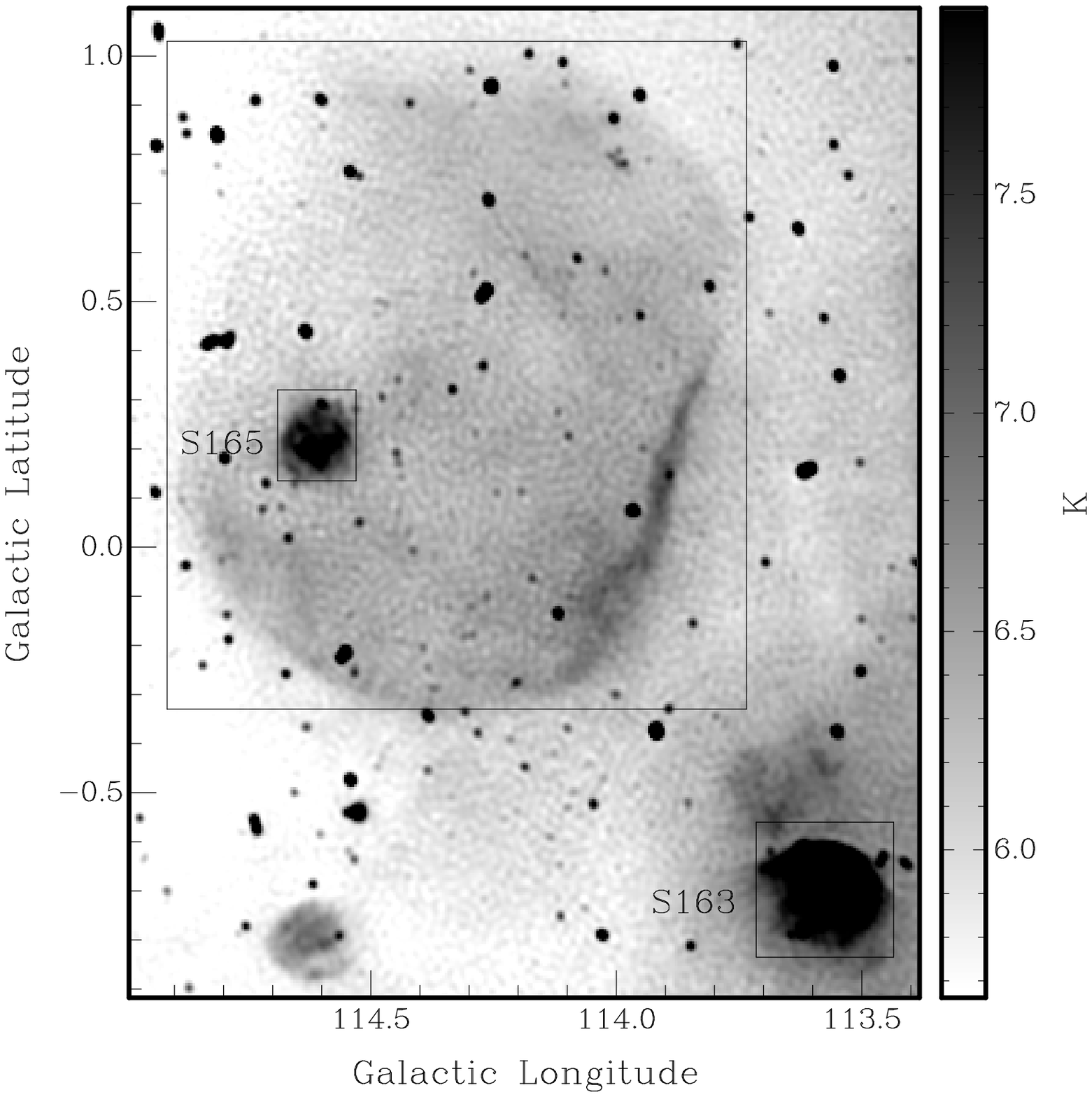}}
\put(-20,100){\includegraphics{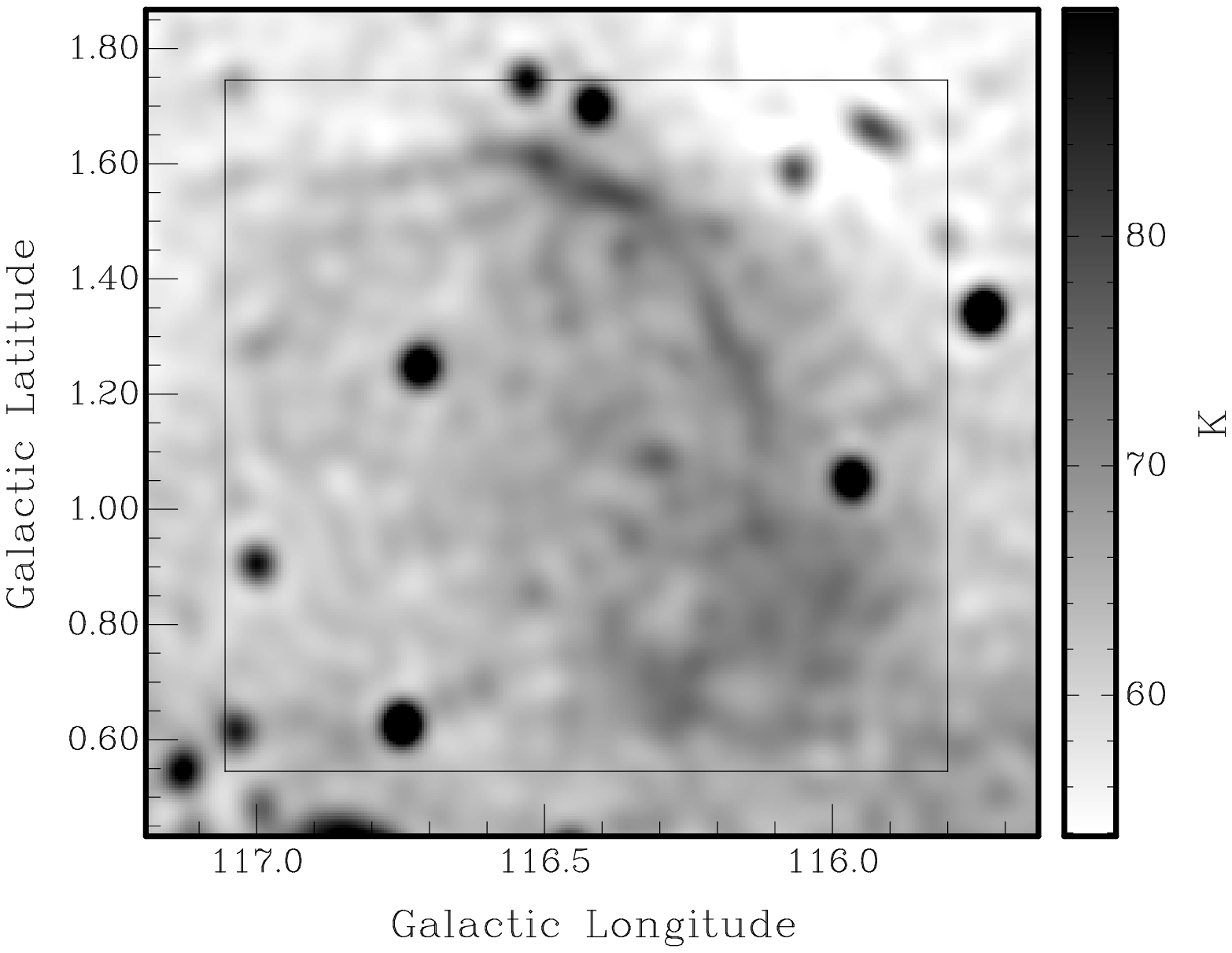}}
\put(190,95){\includegraphics{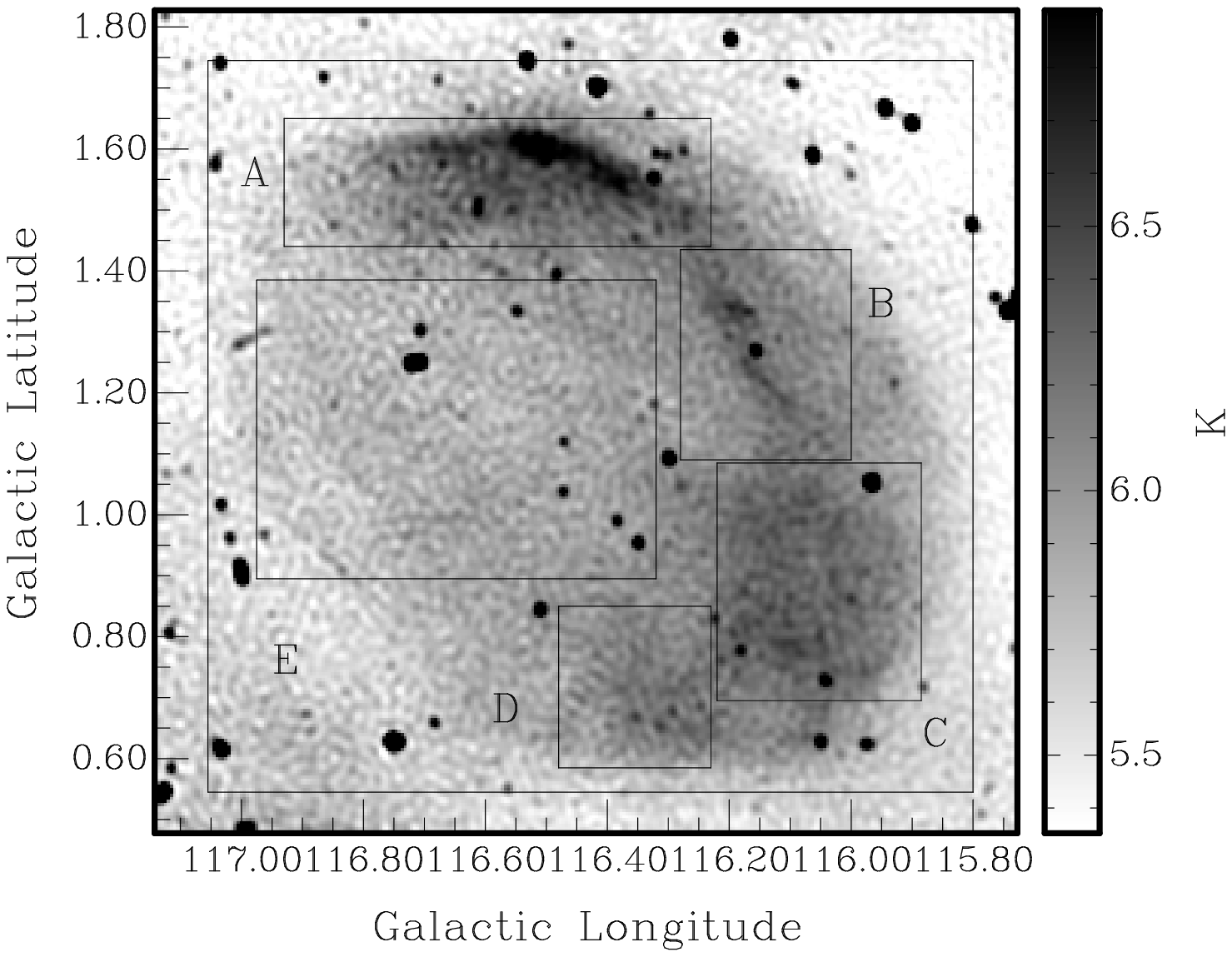}}
\put(-30,-110){\includegraphics{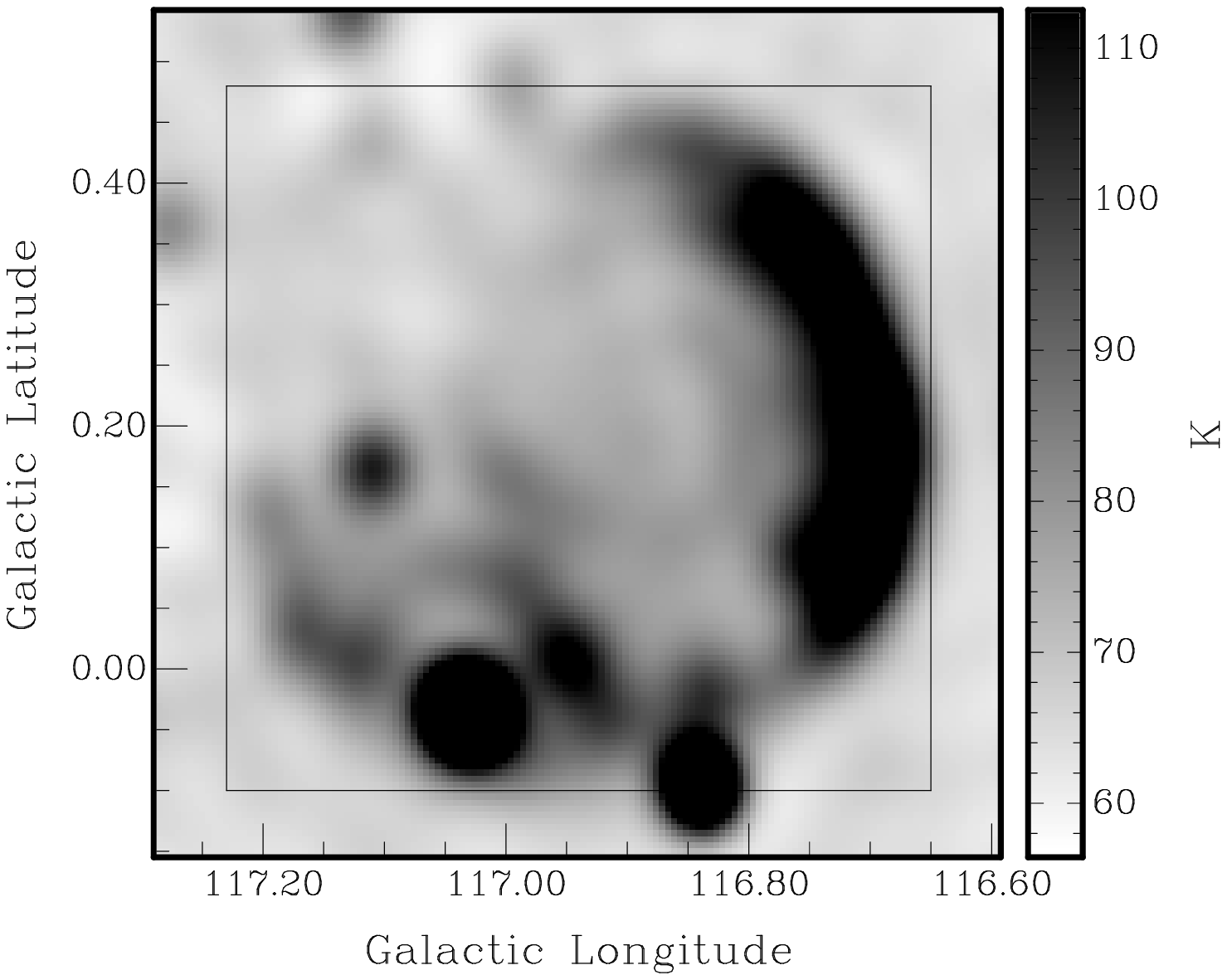}}
\put(190,-110){\includegraphics{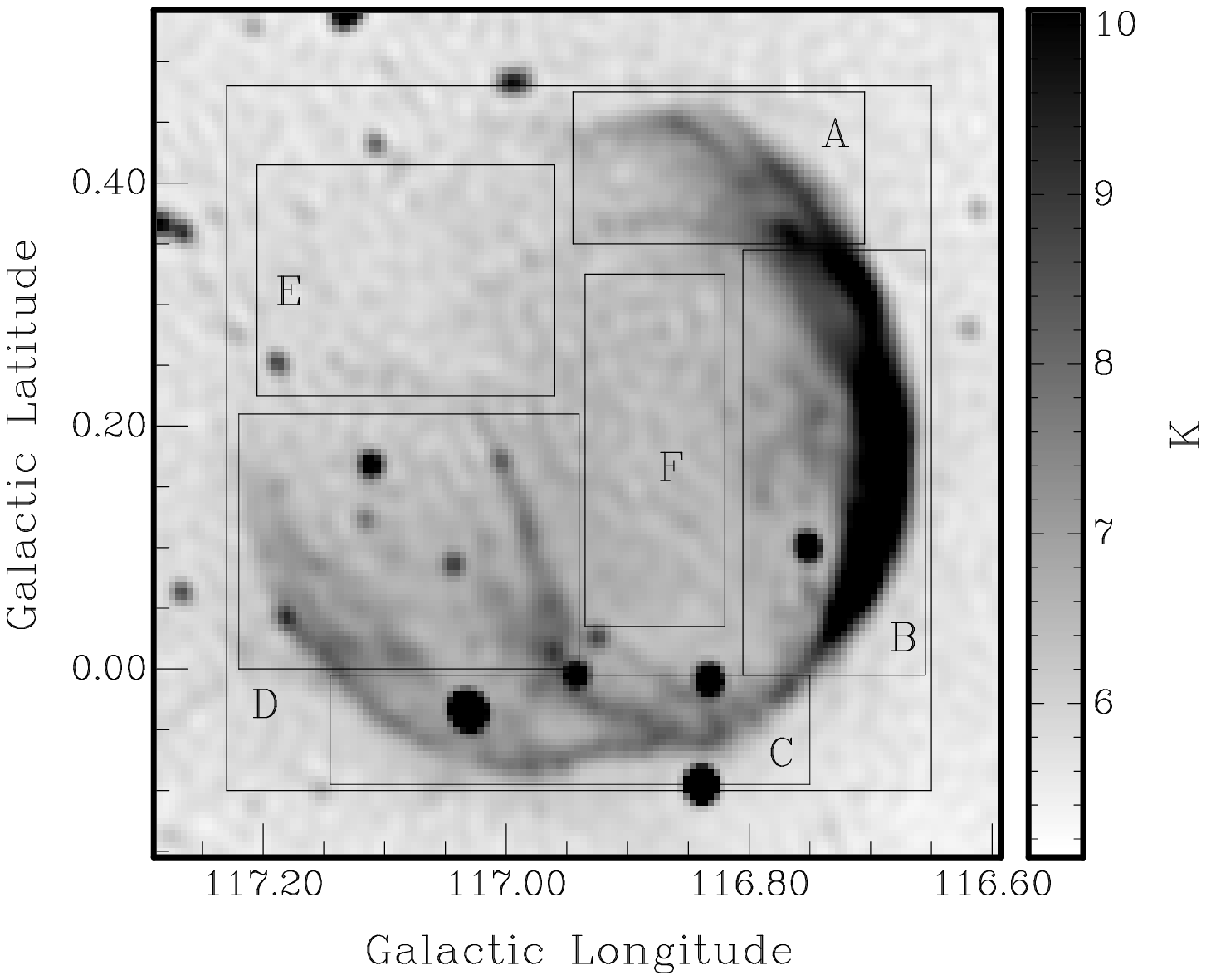}}
\end{picture}
\caption[xx]{The first row of images shows G114.3+0.3 at 408 MHz (left) and 1420 MHz (right). The second row is for G116.5+1.1, and the third for G116.9+0.2. The large boxes used for whole SNR T-T plots are shown in the 408 MHz images. The smaller boxes, labeled with letters and used for sub-area T-T-plots, are shown in the 1420 MHz images.}
\end{figure*}

\begin{figure*}
\vspace{50mm}
\begin{picture}(300,100)
\put(0,-210){\includegraphics{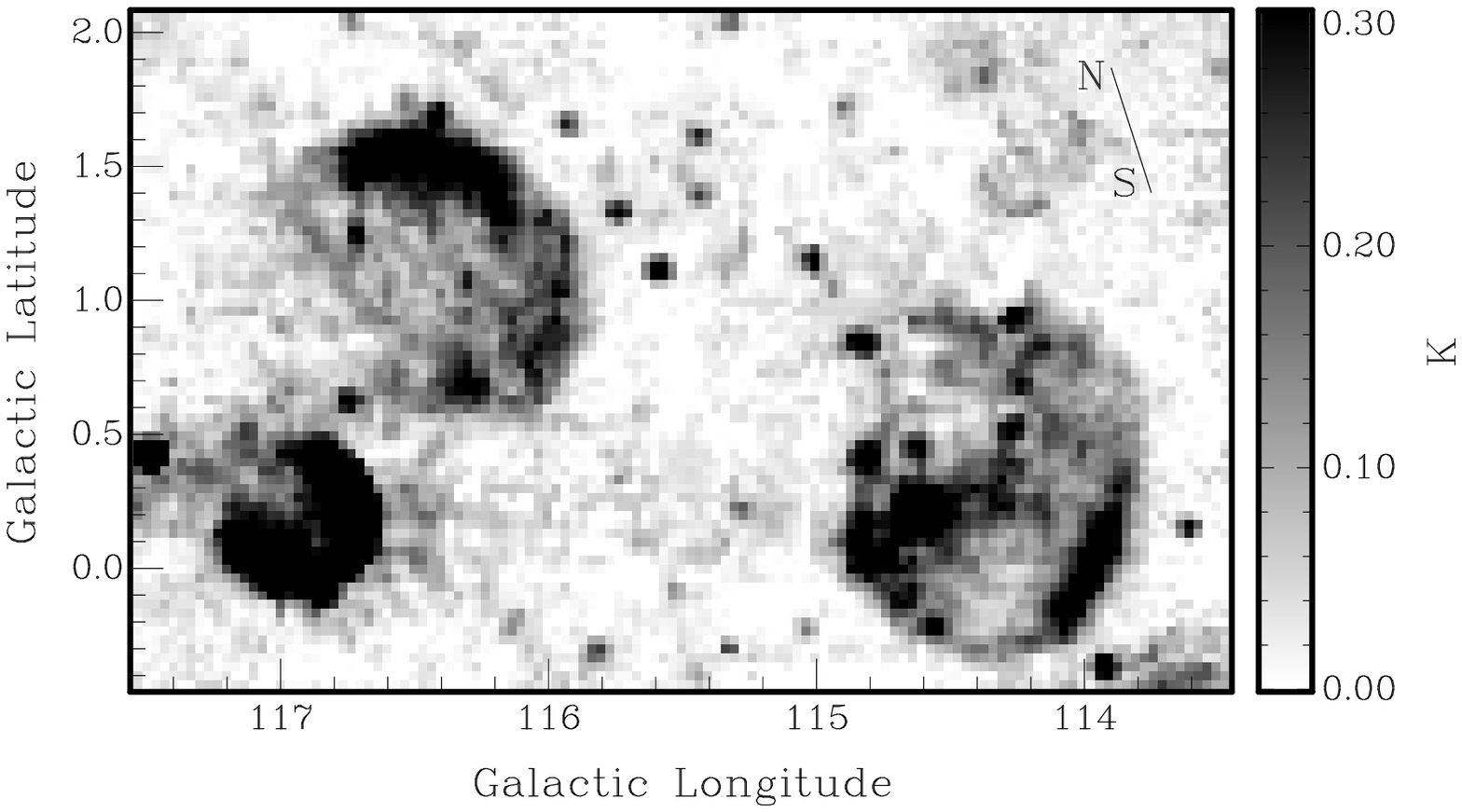}}
\end{picture}
\caption[xx]{The 2695 MHz Effelsberg map of the three supernova remnants, G114.3+0.3, G116.5+1.1 and G116.9+0.2 (CTB 1). The direction of North (N) and South (S) is marked on the upper right.}
\end{figure*}

\subsection{Structure at 408 MHz and 1420 MHz}

The CGPS images at 408 MHz and 1420 MHz are shown in Fig. 1. 
The left panels show the 408 MHz maps, the right panels show the 1420 MHz maps. For reference we also reproduce the 2695 MHz Effelsberg map of the three SNRs in Fig. 2 
(from the Effelsberg public database, F\"urst et al., 1990). 
The Effelsberg map has a resolution of 4.3'. 
The 408 MHz images are seen here for the first time. G114.3+0.3 (Fig. 1 top
panels) is only
faintly visible at 408 MHz, but 408 MHz emission is clearly seen from S165 and the bright 1420 MHz
southeast limb. Comparing with the 1420 MHz image, faint 408 MHz emission from
G114.3+0.3 is seen along the northwest limb and the southeast interior. The strong 408 
MHz background, especially in the southern half of the image, makes G114.3+0.4 difficult
to see except by comparison with the 1420 MHz structure.
There is a bright emission region at l=113.6, b=-0.7, which is the HII region
S163 (Lockman, 1989).

G116.5+1.1 (Fig. 1 middle panels) is more clearly visible at 408 MHz. It shows essentially the same features
as at 1420 MHz with bright north to northeast limb, bright southern interior, and faint 
west to southwest limb. For both G114.3+0.3 and G116.5+1.1 bright compact sources 
contribute significantly to the flux density (see especially the 1420 MHz map for the
number of compact sources across the face of these SNRs). At lower spatial resolution
the compact sources would be confused with the diffuse spatial emission of these SNRs.

G116.9+0.2 (CTB 1, Fig. 1 lower panels) is a bright SNR and has clearly defined shell
structure at 408 MHz which is fully consistent with the 1420 MHz structure convolved
to the lower resolution of the 408 MHz data. Compact sources contribute to the flux
density mainly along the southern limb and the southern interior filament.

\begin{table}
\begin{center}
\caption{408-1420 MHz T-T plot spectral indices
with and without Compact Sources(CS)}
\setlength{\tabcolsep}{1mm}
\begin{tabular}{ccc}
\hline
\hline
     &$\alpha$ & $\alpha$ \\
 Area &including CS &  CS removed \\
\hline
\hline
 All G114.3+0.3& 0.64$\pm$0.34*& 0.68$\pm$0.48\\
\hline
For G116.5+1.1\\
\hline
 A& 0.26$\pm$0.19&0.26$\pm$0.19\\
 B& 0.41$\pm$0.20&0.40$\pm$0.27\\
 C& 0.23$\pm$0.35&0.16$\pm$0.32\\
 D& 0.56$\pm$0.23&0.56$\pm$0.23\\
 E& 0.55$\pm$0.20&0.48$\pm$0.21\\
\hline
All G116.5+1.1&0.34$\pm$0.15&0.28$\pm$0.15\\
\hline
For G116.9+0.2\\
\hline
 A& 0.55$\pm$0.04&0.55$\pm$0.04\\
 B& 0.50$\pm$0.04&0.50$\pm$0.04\\
 C& 0.76$\pm$0.04&0.38$\pm$0.04\\
 D& 0.57$\pm$0.08&0.45$\pm$0.04\\
 E& 0.11$\pm$0.31&0.11$\pm$0.31\\
 F& 0.63$\pm$0.05&0.75$\pm$0.14\\
\hline
 All G116.9+0.4& 0.51$\pm$0.04 & 0.48$\pm$0.04\\
\hline
 \hline
*includes S165.\\
\end{tabular}
\end{center}
\end{table}

\begin{table*}
\begin{center}
\caption{Integrated flux densities and spectral indices of G114.3+0.3 (G114), G116.5+1.1 (G116.5), G116.9+0.2 (CTB 1), and Compact Sources (CS) within three SNRs}
\setlength{\tabcolsep}{1mm}
\begin{tabular}{cccccccccc}
\hline
\hline
Freq.& G114& G114's CS*&G114+CS*& G116.5 & G116.5's CS & G116.5+CS&CTB 1&CTB 1's CS& CTB 1+CS\\
\hline
MHz &Jy&Jy &Jy&Jy&Jy&Jy&Jy&Jy&Jy\\
\hline
\hline
 408 & 12.0$\pm$6.0 & 3.75$\pm$0.24 & 15.8$\pm$6.2 &  15.0$\pm$1.5 & 2.71$\pm$0.15 & 17.7$\pm$1.7  &   15.0$\pm$1.5 & 2.05$\pm$0.13 & 17.1$\pm$1.6\\
 1420&  9.8$\pm$0.8 & 2.08$\pm$0.09  & 11.5$\pm$0.9 &  10.6$\pm$0.6 & 0.79$\pm$0.03  & 11.4$\pm$0.6  &    8.1$\pm$0.4 & 0.73$\pm$0.05  & 8.8$\pm$0.5 \\
\hline
\hline
$\alpha$ &0.16$\pm$0.41&0.47$\pm$0.06 &0.25$\pm$0.32  & 0.28$\pm$0.09&0.99$\pm$0.05 &0.35$\pm$0.09 & 0.49$\pm$0.09 &0.83$\pm$0.07 &0.50$\pm$0.07\\
\hline
\hline
\end{tabular}
\end{center}
* includes S165
\end{table*}

\begin{table}
\begin{center}
\caption{Integrated Flux Densities (FD) of G114.3+0.3}
\setlength{\tabcolsep}{1mm}
\begin{tabular}{ccccc}
\hline
\hline
Freq. &Beamwidth&FD & references\\
MHz   &arcmin   & Jy               & \\
\hline
\hline
 102.0 & 48$\times$25 & 13.0$\pm$3.0 & 1994, Kovalenko et al.\\
 408.0 &3.4$\times$3.9& 12.0$\pm$6.0* & this paper\\
 1420.0& 9$\times$9 & 5.3$\pm$0.8 & 1981, Reich\&Braunfurth\\
 1420.0& 1$\times$1.13 & 9.8$\pm$0.8* & this paper\\
 2695 & 4.3$\times$4.3& 8.9$\pm$1.2 & see text \\
 2700.0&4.6 $\times$4.6   & 4.4$\pm$0.4 & 1981, Reich\&Braunfurth \\
 \hline
\hline
\end{tabular}

\caption{Integrated flux densities (FD) of G116.5+1.1}
\setlength{\tabcolsep}{1mm}
\begin{tabular}{ccccc}
\hline
\hline
Freq. &Beamwidth &FD & references\\
MHz   &arcmin    & Jy               & \\
\hline
\hline
 102.0  &48$\times$25 & 77.0 $\pm$11.0& 1994,Kovalenko et al.\\
 408.0  & 3.4$\times$3.8  & 15.0$\pm$1.5* & this paper\\
 1420.0 &  9 $\times$9    &  8.0$\pm$0.8 & 1981,Reich\&Braunfurth\\
 1420.0 &  1 $\times$1.12 & 10.6$\pm$0.6* & this paper\\
 2695 & 4.3$\times$4.3& 6.3$\pm$1.1 & see text \\
 2700.0 & 4.6$\times$4.6  &  4.7 $\pm$0.4 & 1981,Reich\&Braunfurth \\
\hline
\hline
\end{tabular}

\caption{Integrated flux densities (FD) of G116.9+0.2}
\setlength{\tabcolsep}{1mm}
\begin{tabular}{ccccc}
\hline
\hline
Freq. &Beamwidth &FD & references\\
MHz   &arcmin    & Jy               & \\
\hline
\hline
 102.0 &48 $\times$25 & 51.0$\pm$10.0 & 1994,  Kovalenko et al.\\
 408.0 &3.4$\times$3.8&  15.0$\pm$1.5* & this paper\\
 610.0& 16$\times$20 & 10.9$\pm$1.0 & 1971,1973, Willis\&Dickel\\
 1400.0& 9.7$\times$9.7 & 9.4$\pm$1.5 & 1971,1973, Willis\&Dickel\\
 1420.0& 9$\times$9 &7.8$\pm$0.8(7.1) & 1981, Reich\&Braunfurth \\
 1420.0& 2$\times$2 &8.3$\pm$0.5(7.6) & 1982, Landecker et al. \\
 1420.0& 1$\times$1.13 & 8.1$\pm$0.4*& this paper\\
 1667.0& 22$\times$22& 5.5$\pm$2.0 & 1973, Willis, A.G.\\
 2695 & 4.3$\times$4.3& 6.1$\pm$1.0 & see text \\
 2700.0& 4.6$\times$4.6& 4.8$\pm$0.4(4.45)& 1981, Reich\&Braunfurth\\
 2740.0& 5$\times$5 & 4.2$\pm$0.4 (3.85)& 1974, Velusamy\&Kundu\\
 5000.0& 6.8$\times$6.8 & 3.0$\pm$0.3 (2.8)& 1977, Angerhofer et al. \\
\hline
\hline
\end{tabular}
\end{center}
 The value in () means compact sources' contributions have been subtracted from the 
SNR's flux density by using 610 and 1410 MHz fluxes and its spectral index of 
compacts sources (Dickel and Willis, 1980).

*The flux density excludes compact sources. 
\end{table}

\subsection{T-T plot Spectral Indices}

Bright compact sources affect both integrated flux density for the SNRs and 
spectral indices, so we correct for the effects of compact sources below. 
First we discuss spectral indices between 408 MHz and 1420 MHz based on 
the T-T plot method (for a description of the T-T plot method see e.g. 
Leahy and Roger, 1998).
For the T-T plot analysis, first a single region for the whole of each SNR is used, 
as shown in Fig. 1. The SNRs yield the T-T plots shown in Fig. 3 (G114.3+0.3 left;
G116.5+1.1 middle; G116.9+0.2 right). 
Two cases are considered: using all pixels including compact sources; and 
excluding compact sources. 
The compact sources' influence on the T-T plots is clearly seen in Fig. 1.  
When we quote spectral index errors, we combine the formal error from the fit to the
T-T plot with the spectral index error of $\simeq0.03$ arising from the calibration 
errors of the 408 and 1420 MHz flux densities. In most cases, the fit error dominates.

\begin{figure*}
\vspace{70mm}
\begin{picture}(60,160)
\put(-27,138){\includegraphics{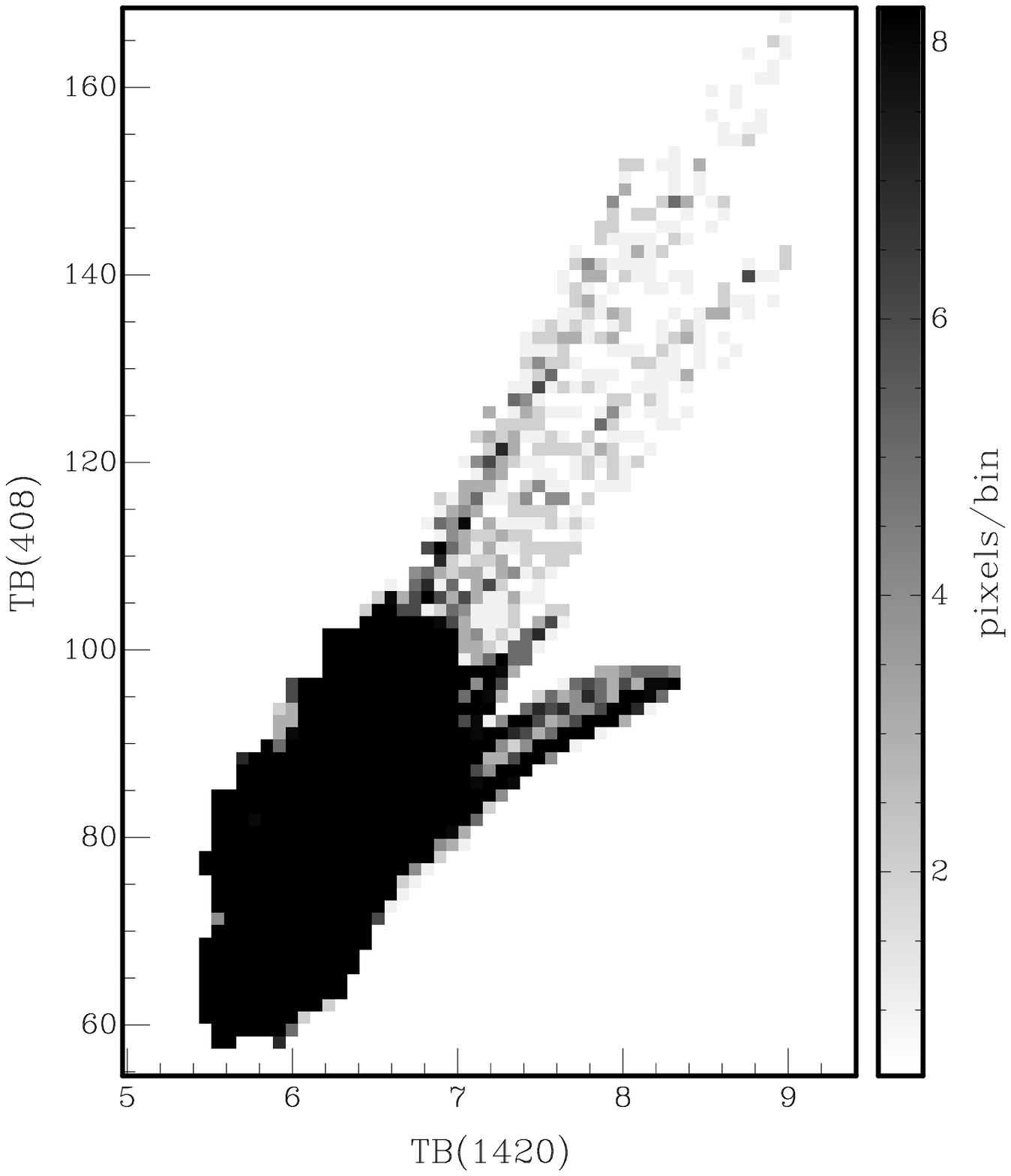}} 
\put(120,140){\includegraphics{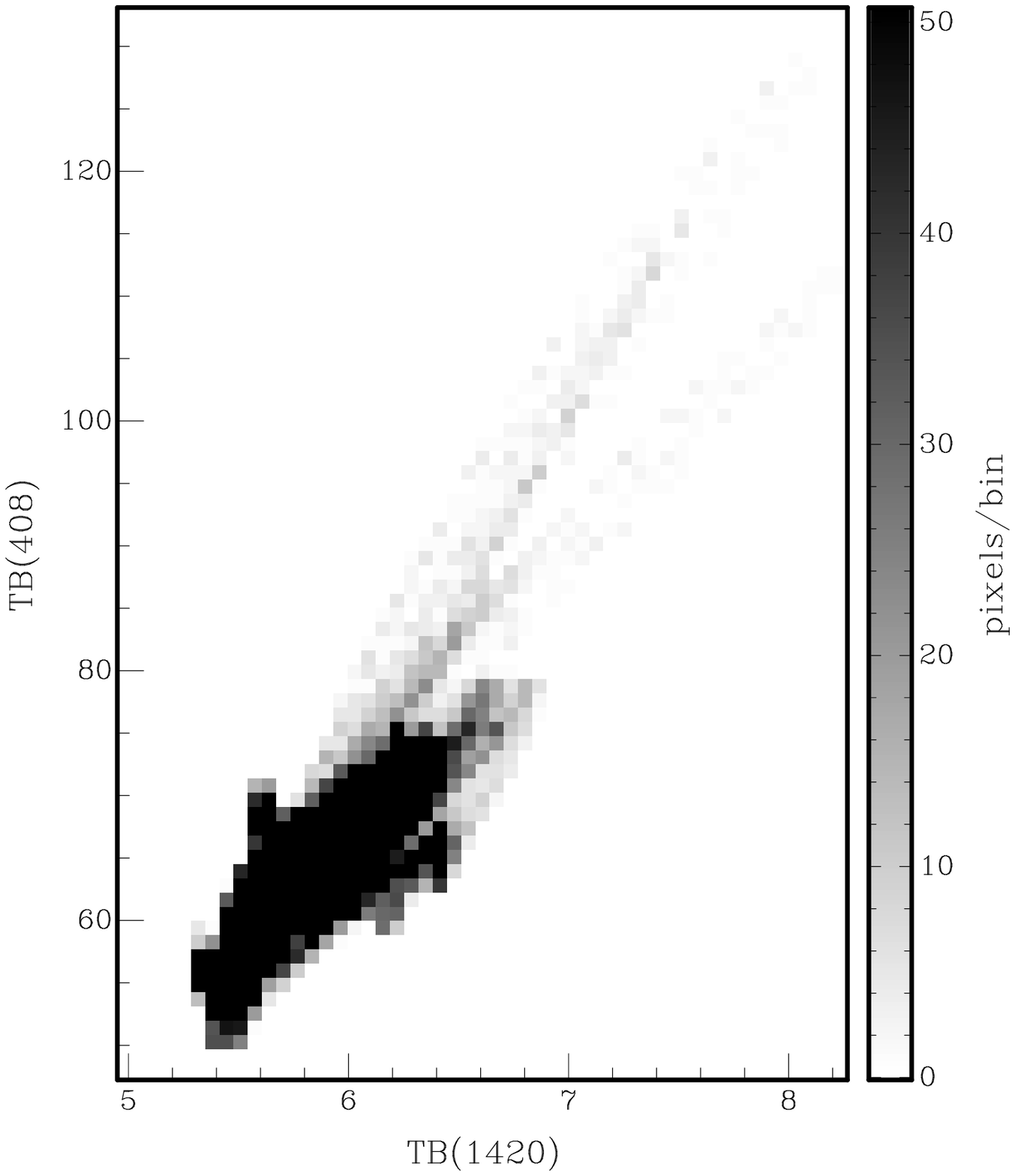}} 
\put(270,138){\includegraphics{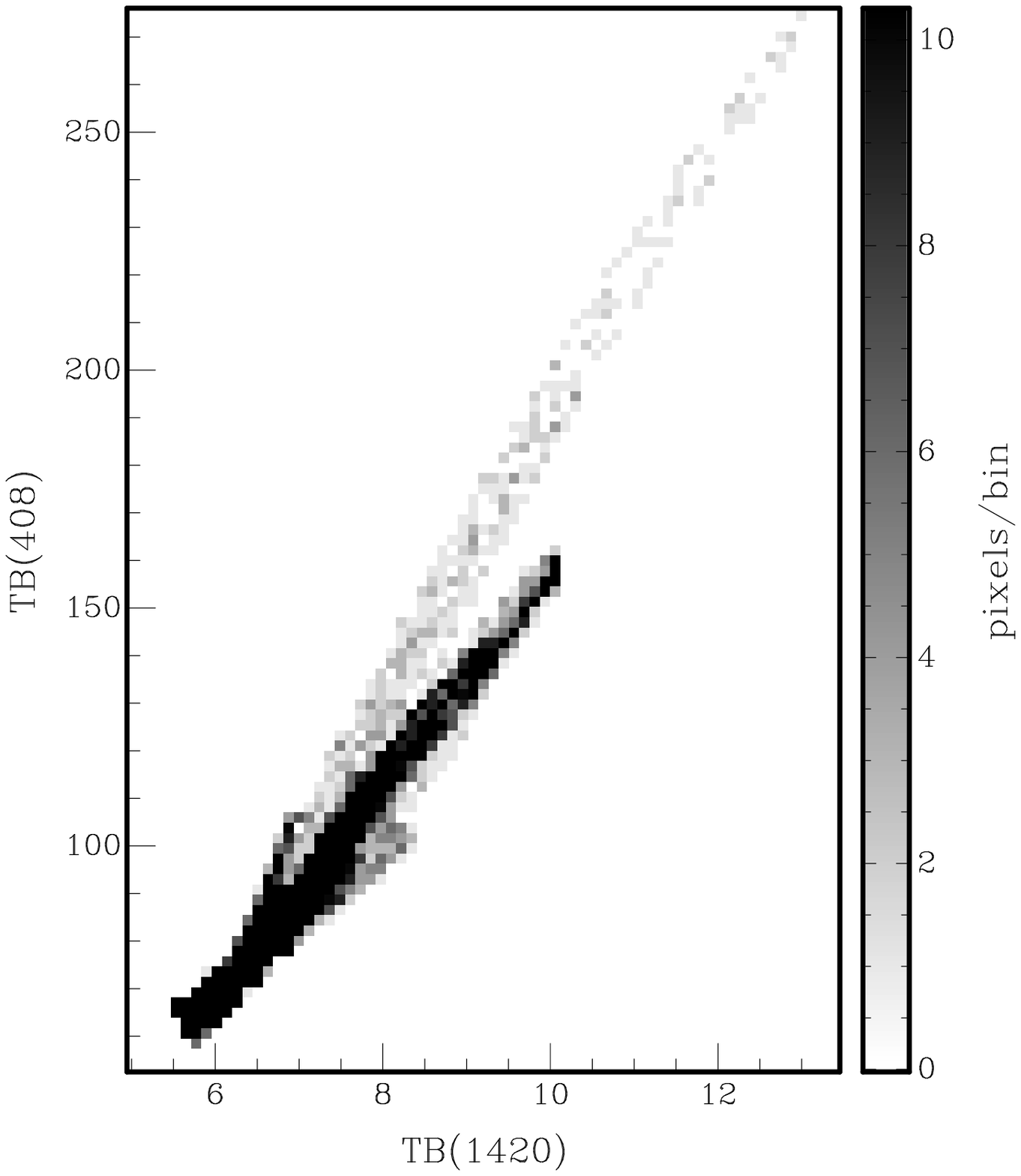}}
\put(-35,-45){\includegraphics{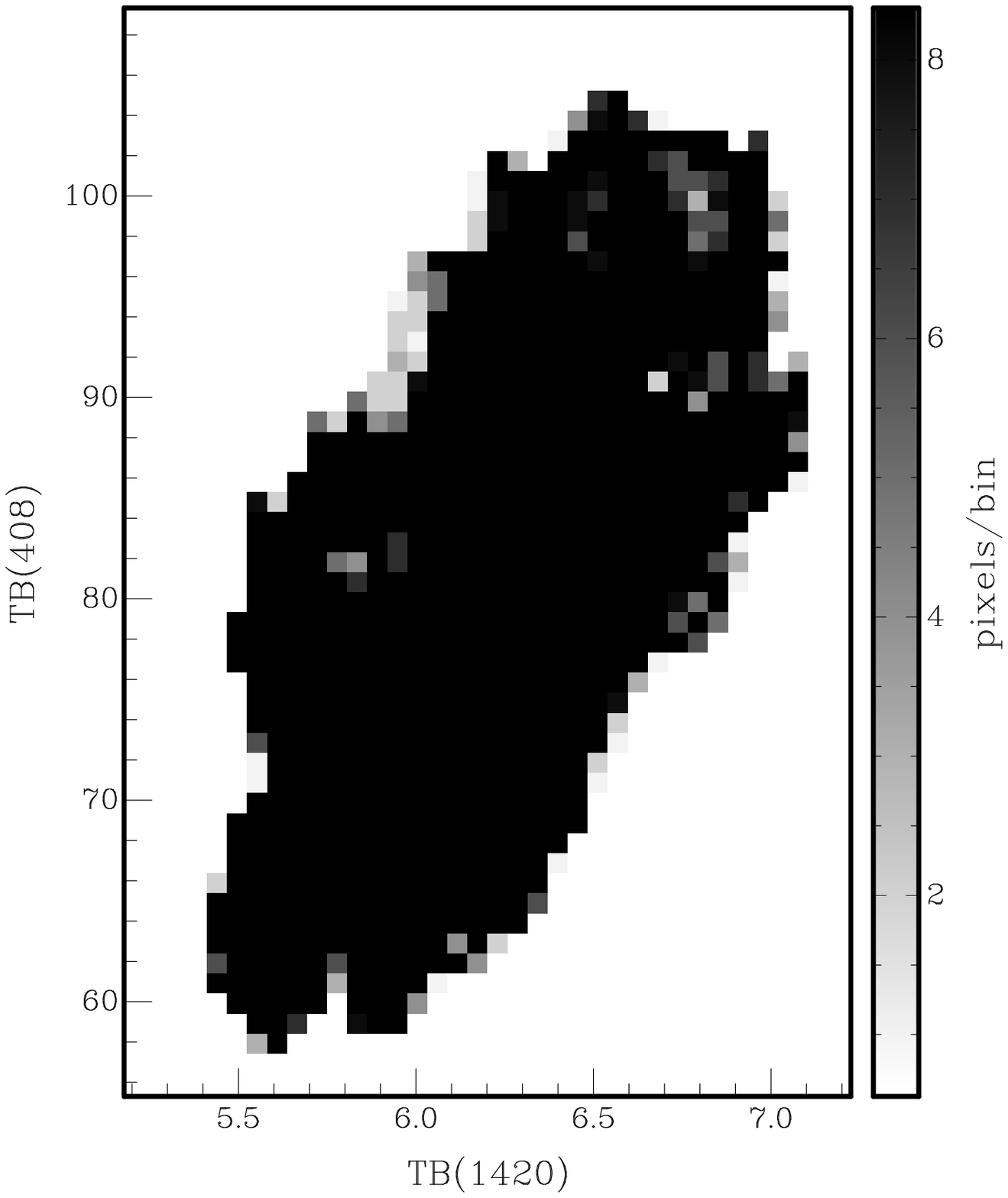}}
\put(120,-45){\includegraphics{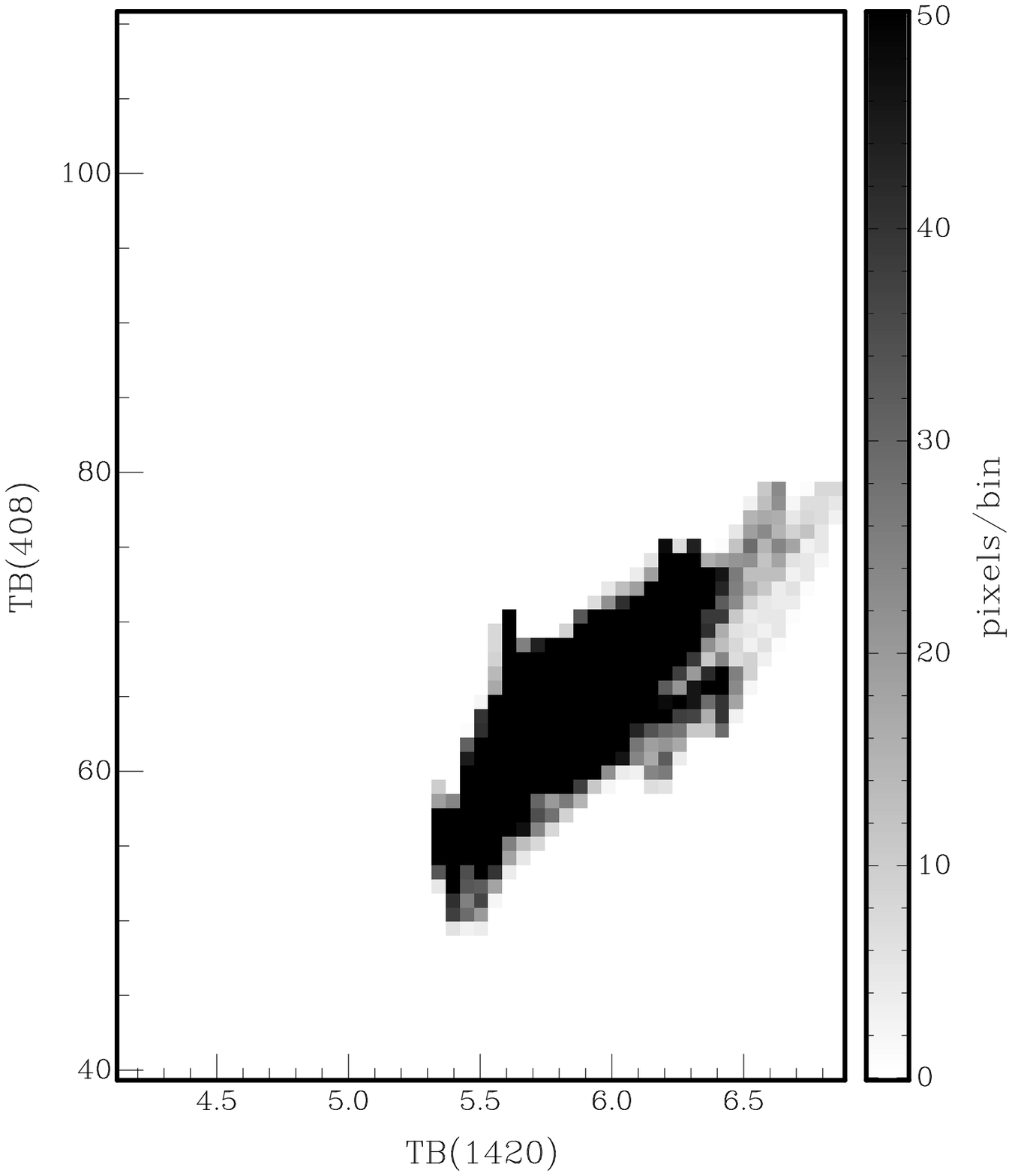}}
\put(280,-45){\includegraphics{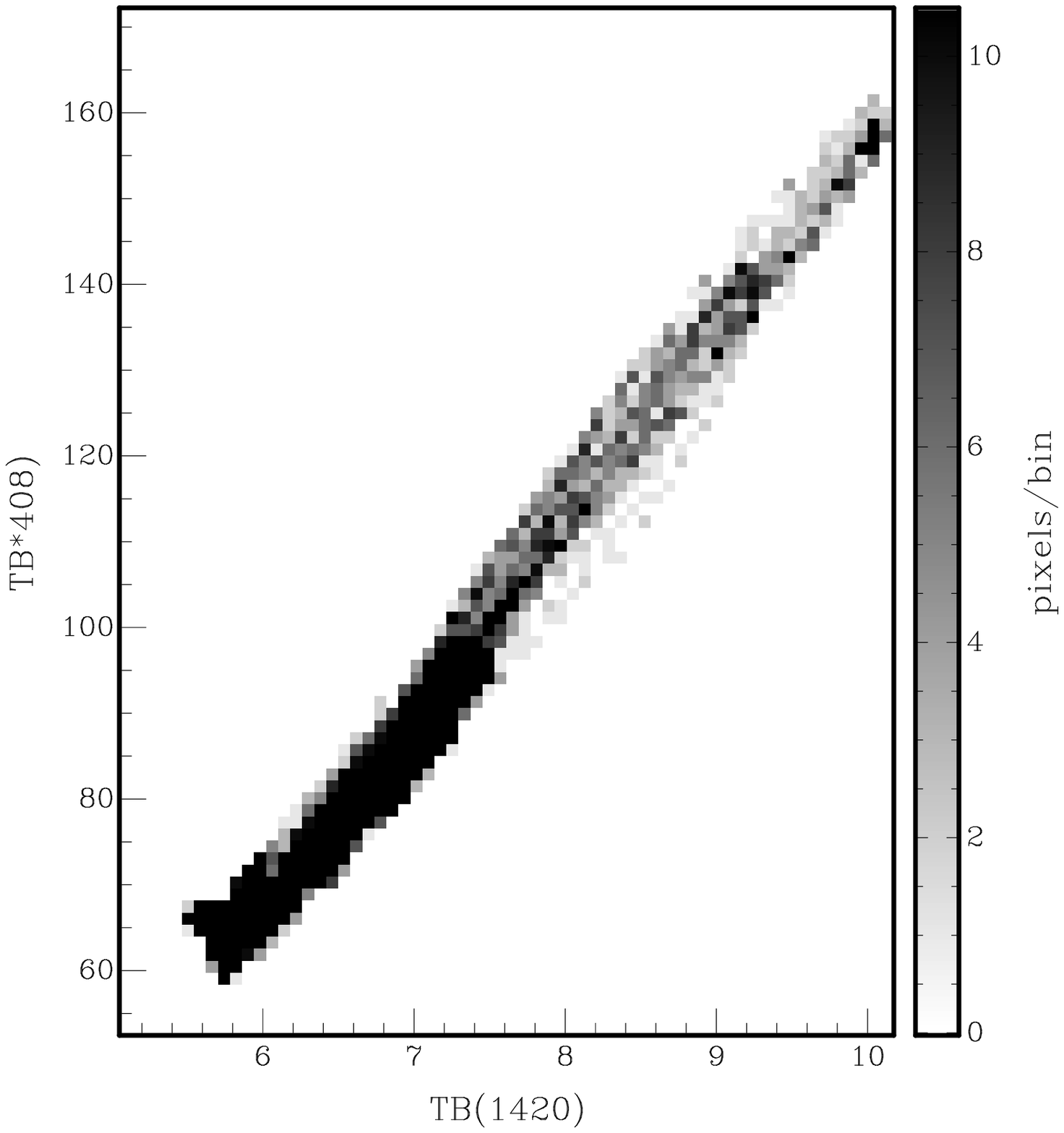}}
\end{picture}
\caption[xx]{Whole SNR 408-1420 MHz T-T plots for G114.3+0.3 (left), G116.5+1.1(center), and G116.9+0.2(right). From top to buttom for G114.3+0.3: plot for map including compact sources ($\alpha$=0.64 $\pm$0.34); 
plot for compact sources removed from analysis ($\alpha$=0.68$\pm$0.48). The respective values of $\alpha$ for G116.5+1.1 are 0.34$\pm$0.15 and 0.28$\pm$0.15; for G116.9+0.2, 0.51$\pm$0.04 and 0.48$\pm$0.04.}
\end{figure*}

Next, the SNRs are subdivided into smaller areas which are labeled in Fig. 1 to search for spatial variations in spectral index.  
Table 1 lists the results for both cases of analysis: 
including compact sources and removing compact 
sources. There is not a large difference in results between the two methods,
despite the contamination of compact sources for the first method. 
By inspecting the T-T plots, as in Fig. 3, the agreement is just a coincidence 
caused by a mixture of steep and flat spectral indices for
compact sources which does not result in a large change in spectral index.
For our other studies of spectral indices of SNRs (Tian and Leahy,
2005; Leahy and Tian, 2005; Tian and Leahy, 2006), the compact sources generally
had significantly steeper spectral indices than the SNRs, so the correct analysis
did produce different spectral indices.
From now on we discuss
spectral indices derived with compact sources removed, unless specified otherwise.
We also carried out a T-T plot analysis for S163
and we verify it has a thermal spectral index similar to S165. 

\subsection{Integrated Flux Densities and Spectral Indices}

From the 
408 MHz and 1420 MHz maps we have derived integrated flux densities for the three SNRs with diffuse background subtracted.
The resulting 408 MHz to 1420 MHz spectral indices, using flux densities without compact sources, 
are 0.16$\pm$0.41 for G114.3+0.3,  0.28$\pm$0.09 for G116.5+1.1, and 0.49$\pm$0.09 for G116.9+0.2.  
Table 2 lists the flux densities and spectral indices for the SNRs and the
compact sources within each SNR. 

\begin{figure*}
\vspace{40mm}
\begin{picture}(0,70)
\put(-20,-20){\includegraphics{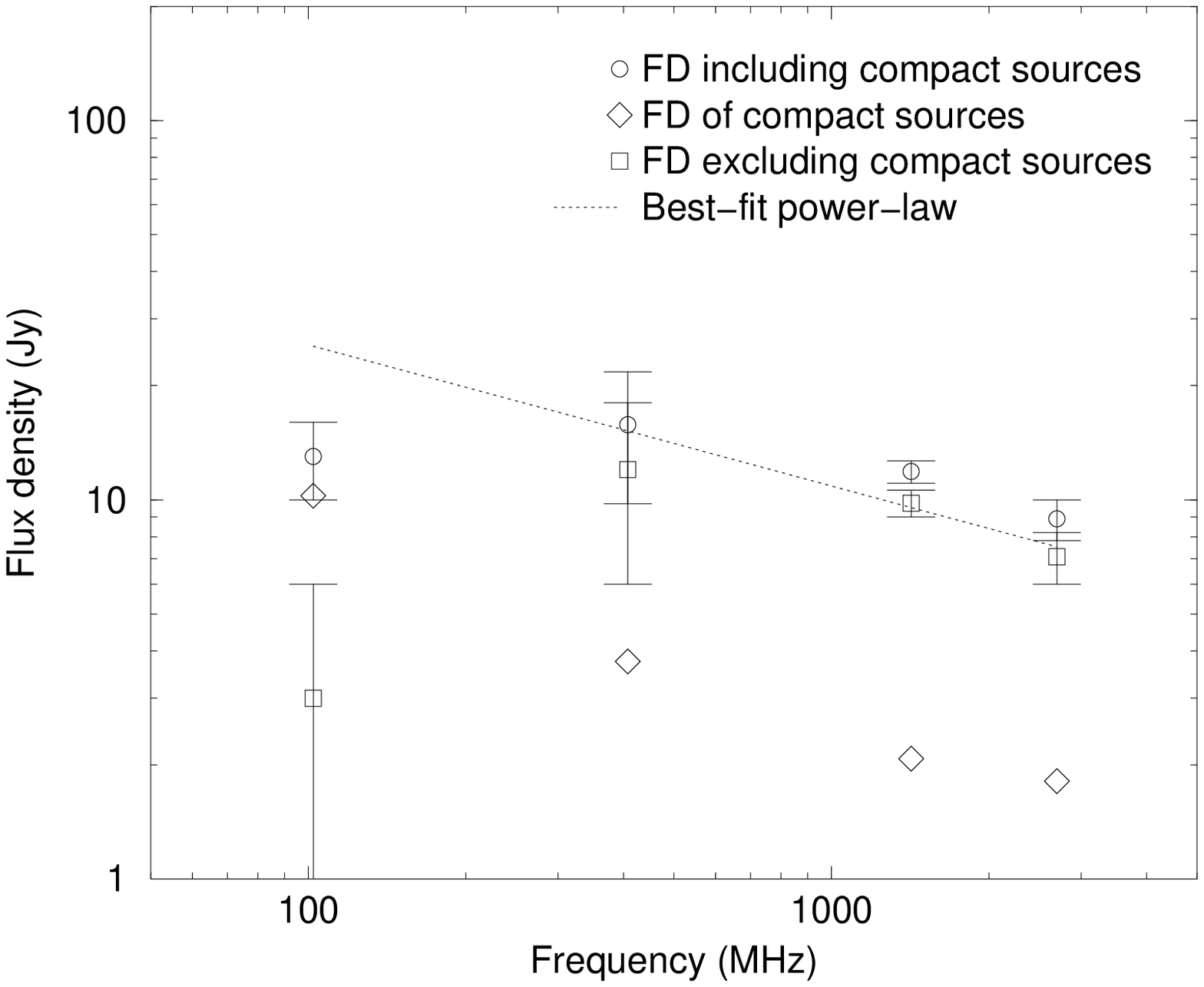}}
\put(132,-20){\includegraphics{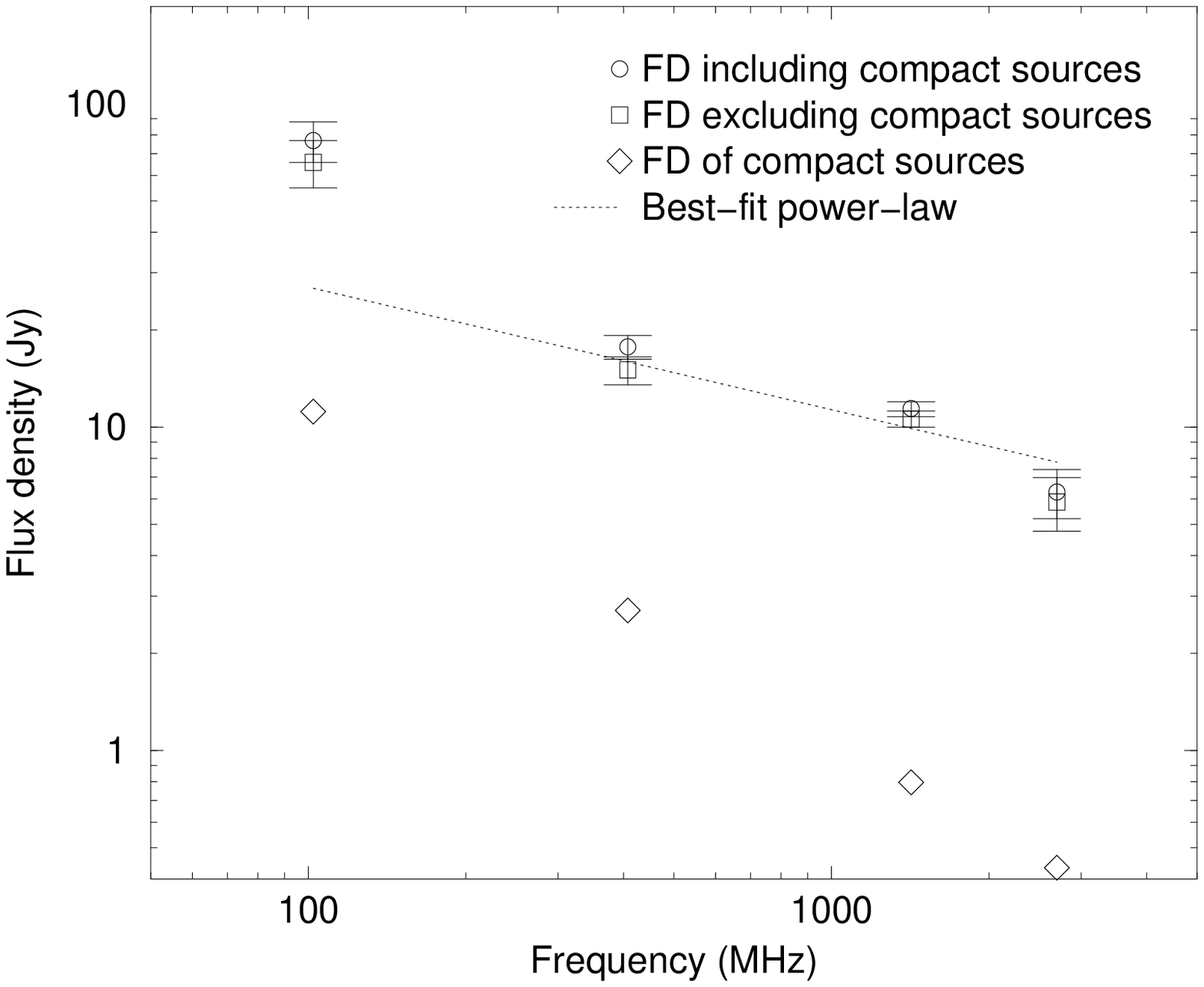}}
\put(285,-20){\includegraphics{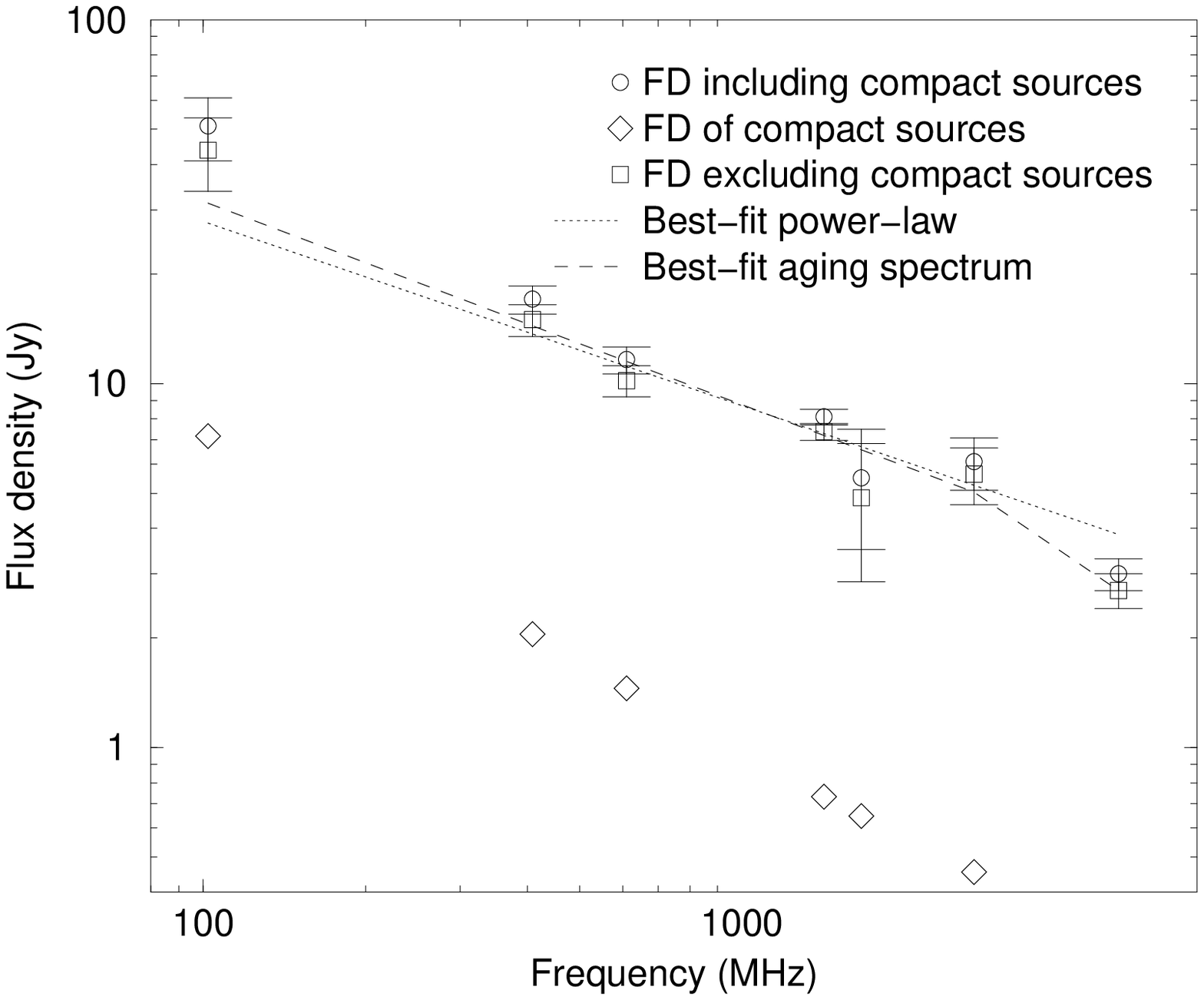}}
\end{picture}
\caption[xx]{Radio spectrum of G114.3+0.3 (left), G116.5+1.1 (middle) and G116.9+0.2 (right).  G114.3+0.3 has a best-fit spectral index $\alpha$=0.37, G116.5+1.1 has best fit $\alpha$=0.38  and G116.9+0.2 has $\alpha$=0.51 (see text for details).} 
\end{figure*}

\section{Discussion}
\label{sect:Dis}
\subsection{G114.3+0.3}

Published integrated flux densities for G114.3+0.3  
at other frequencies are given in Table 3. Fig. 4 (left) shows 
flux densities before and after correcting for the flux density of compact
sources. 
We carry out all spectral fits here for all three SNRs using fluxes with compact
sources removed.
The best fit power-law spectrum to all four flux-densities gives a spectral index 
of -0.1 and $\chi^2$=8. The 102 MHz point is well below the power law fit. 
Referring to Fig.1, one can see a strong
steep spectrum background in the southern half of the G114.3+0.3 408 MHz image. 
Also at the higher spatial resolution of the 1420 MHz map, there is a low
background just outside the south and west limb, but it rises to high values
further outside the limb. If one viewed G114.3+0.3 at low spatial resolution, 
such as that for the 102 MHz observation, 
one would see the bright emission south and west of G114.3+0.3 smeared into
the lower background just outside the limb. This would cause one to 
overestimate the background compared to the true background. 
Thus the flux density at 102 MHz should be higher than the published value. 
A spectral fit to the data without the 102 MHz point gives a spectral index of
0.37 (90$\%$ uncertainty range 0.05 to 0.66). This spectral fit is shown in Fig. 4 (left).
This value is consistent with the 408-1420 MHz spectral index from the T-T plot of 0.7$\pm0.5$
and from 408 and 1420 MHz integrated fluxes (0.16$\pm$0.41). 
The various spectral index measurements all have large uncertainties, due to
the faintness of G114.3+0.3 and the high background in the region.


\subsection{G116.5+1.1}

The integrated flux densities for G116.5+1.1 are given in Table 4.
The best fit $\alpha$ to all four data points is 0.48 with
$\chi^2$=15.  The 102 MHz point is well above the best fit power-law. The
spectral fit to the data, omitting the 102 MHz point, gives $\alpha$=0.38 ($\chi^2$=4.7) 
and 90$\%$ uncertainty range 0.25 to 0.50.  
This value is consistent with the 408-1420 MHz spectral index from the T-T plot of 0.28$\pm0.15$
and from 408 and 1420 MHz integrated fluxes of 0.28$\pm$0.09.
Fig. 4. (middle) shows the radio spectrum with correction for compact sources
and the best fit power law spectrum ($\alpha$=0.38). 
Table 1 shows the T-T plot spectral indices for subregions of G116.5+1.1. They are all
consistent within errors with the whole SNR T-T plot value of 0.28$\pm0.15$. So there is
no evidence for spectral index variations in G116.5+1.1 to the level of $\simeq0.2$ in
spectral index.

Mavromatakis et al. (2005) image G116.5+1.1 in narrow band optical filters and obtain optical 
spectra of several bright filaments inside and northeast of the radio SNR. These yield
line ratios indicative of shock heated gas which extends to the northeast of
G116.5+1.1 by about 20 arcmin. We have studied the HI maps from the CGPS. 
They have the clearest
association with the outline of the radio SNR in the range -41 to -45 km/s. 
That HI map (their Fig.4), interestingly, has a gap in HI just where the 
shocked optical filaments extend beyond the boundary of the radio SNR. 
This suggests that the optical filaments are a breakout region from G116.5+1.1
and not a separate SNR.
From the HI velocity Mavromatakis et al. (2005) propose that G116.5+1.1 is in
the Perseus arm. However, this contradicts the conclusion from Yar-Uyaniker et al. (2004)
that it is a local arm SNR, based on detection of polarized 
radio emission.
Neither argument is conclusive but the distance from Yar-Uyaniker et al. is 
probably better, due to the confusion limited
nature of the HI emission in the galactic plane.  

\subsection{G116.9+0.2 (CTB 1)}

Published integrated flux densities for CTB 1
are given in Table 5, and shown in Fig. 4 (right). For CTB 1 the correction
for compact source flux density is less important than for the other two SNRs, though
the correction is important for spectral indices derived from subareas C and D (Fig.1).
The best fit spectral index to all 7 data points is 0.65, with $\chi^2$=8.
The 102 MHz point is above the power-law fit, and the 5 GHz point is below it.
Fitting without the 102 MHz and 5 GHz points gives a power-law which agrees with all
of the remaining data points, with $\alpha$=0.51 ($\chi^2$=2.8) and
90$\%$ range 0.37 to 0.64. This fit is shown in Fig. 4 (right). 
Next we carry out a fit including the 102 MHz and 5 GHz points, 
but use a synchrotron aging model. This gives $\alpha$=0.56 ($\chi^2$=4.8), a
90$\%$ range 0.29 to 0.63, and a turnover frequency of 2.9 GHz. 

The different methods for determining the spectral index of CTB 1 all agree and yield a value
consistent with the best determined value of 0.48$\pm0.01$ from the T-T plot method. 
The spectral fit to multi-frequency flux 
density values gives weak evidence (1.6$\sigma$)
for a spectrum turnover by 0.5 in spectral index at a frequency of about 2.9 GHz. A more accurate 
measurement of the 5 GHz flux density could confirm this result.
Here we show that a turnover at $\nu_b$=2.9 GHz is quite feasible for CTB 1. From  $\nu_b$, we get
the electron break energy $E_b=1.9\times 10^{-9}(B/2\mu G)^{-0.5}$J, where $B$ is the magnetic field
in the synchrotron emitting plasma. An initial power law distribution of electrons
will, over time, have a break in its energy spectrum by 1 in energy index at $E_b$, with $E_b$
increasing with time, t, according to (e.g. Longair, 1981): $E_b=(a_s B^2 t)^{-1}$. $a_s$ is the 
synchrotron energy loss coefficient for energy loss by a single electron: $dE/dt=-a_s B^2 E^2$.
The restriction that the turnover frequency $\nu_b$ gives is a relation between the age of the
supernova remnant and its magnetic field. For $\nu_b$=2.9 GHz, $t=2.7\times10^8(B/2\mu G)^{-3/2}$yr.
So if CTB 1 is $10^4$ yr old, the magnetic field is 1.8 mG. Evidence for high magnetic field and 
high density for CTB 1 is given by the optical observations of Fesen et al. (1997), particularly for
the bright emission along the western limb.

Finally we note evidence for spectral index variations in CTB 1 (see Fig. 1 and Table 1). 
Regions A, B, D, E and F are consistent with a spectral index of 0.50. 
The southern limb (region C) is flatter ($\simeq3\sigma$). This could be caused by variations in the shock causing electron spectrum
variations or by variations in magnetic field causing changes in synchrotron aging. However to 
confirm these results and to study
the matter further, higher spatial resolution observations are required at 408 MHz.

\section{Conclusion}
\label{sect:Con}
New observations of G114.3+0.3, G116.5+1.1, and G116.9+0.2 (CTB 1) at 408 MHz have
been combined with 1420 MHz observations to study the spectral index of these three
SNRs. We remove the effect of compact sources' flux
densities for calculation of spectral indices by the methods of T-T plot and integrated
flux densities. We also use published integrated flux densities and correct these for
compact sources. The resulting spectral indices for the three SNRs are consistent: 
0.37 for G114.3+0.3,  0.38 for G116.5+1.1, 0.51 for CTB 1. The first two have larger uncertainty due to the
faintness of the SNRs and background variations. For CTB 1, the spectral index is very
well determined, perhaps the best yet of any supernova remnant. There is some evidence for
a spectral turnover in the radio spectrum of CTB 1 near 3 GHz, which can be interpreted as
due to synchrotron aging in a strong (mG) magnetic field, consistent with the optical 
observations of Fesen (1997).  

\begin{acknowledgements}
We acknowledge supports from the Natural Sciences and Engineering Research Council of Canada. The DRAO is operated as a national facility by the National Research Council of Canada. 
The Canadian Galactic Plane Survey is a Canadian project with international partners. 
\end{acknowledgements}

\end{document}